%% file: bm.tex
\documentclass[vecphys]{svmult}

\usepackage{makeidx}         
\usepackage{graphicx}        
\usepackage{multicol}        
\usepackage[bottom]{footmisc}

\makeindex             

\begin{document}

\title*{Transonic properties of the accretion disk around compact objects}
\author{Banibrata Mukhopadhyay}
\institute{Astronomy and Astrophysics Programme, Department of Physics,
Indian Institute of Science, Bangalore-560012, India
\texttt{bm@physics.iisc.ernet.in}
}
%
%
\maketitle

An accretion flow is necessarily transonic around a black hole.
However, around a neutron star it may or may not be transonic,
depending on the inner disk boundary conditions influenced by
the neutron star. I will discuss various transonic behavior
of the disk fluid in general relativistic (or pseudo general
relativistic) framework. I will address that there are four types of 
sonic/critical point possible to form in an accretion disk.
It will be shown that how the fluid
properties including location of sonic points vary with angular
momentum of the compact object which controls the overall disk dynamics and outflows. 

\section{Introduction}
\label{intro}

Accretion disks serve to identify compact objects, mostly black holes
and neutron stars, in the universe. The most common way to understand its formation is 
in a binary system where matter is pulled off 
a companion star and settles on to the compact object in the form of a disk. 
As practically black holes and neutron stars cannot be seen, 
by detecting and analyzing light rays off an accretion disk one can understand
the properties of the central compact object.
Other examples of the accretion disk are the protoplanetary
disks, disks around active galactic nuclei, in star-forming systems, and
in quiescent cataclysmic variables etc. 

The molecular viscosity in an accretion disk is rather low. Shakura \& Sunyaev \cite{ss73}
proposed that there is a significant turbulent viscosity explaining transport of matter inwards
and angular momentum outwards with the Keplerian angular momentum profile \cite{pringle}. However, until
the work by Balbus \& Hawley \cite{bh} origin of the instability and plausible turbulence was
not understood. Applying the idea of Magneto-Rotational-Instability (MRI) described by
Velikhov \cite{vel59} and Chandrasekhar \cite{chandra60}, they
showed that the Keplerian disk flow may exhibit unstable modes under perturbation in the 
presence of a weak 
magnetic field. This plausibly generates turbulence which results transport. However, there
are several accretion disk systems where gas is electrically neutral in charge and thus the magnetic field
does not couple with the system. Examples of such system are protoplanetary and star-forming disks, 
disks around active galactic nuclei and in quiescent cataclysmic variables. Hence in these
systems MRI is expected not to work to generate turbulence. While the origin of the 
transport in such
systems is still an ill understood problem, some mechanisms have been proposed by various
groups including the present author \cite{tev,yek,um04,mukh05a,mukh05b,mukh06,zahn}. 

However, the problems of the transport and then the viscosity are not severe if the disk angular
momentum profile deviates from the Keplerian type to sub-Keplerian type. In a sub-Keplerian disk,
the gravitational force dominates over the centrifugal force resulting in a strong advective 
component of the infalling matter \cite{abrazur,c89,ny}. 
Due to the dominance of the gravity, the matter even with a constant
angular momentum may transport inwards easily. An accretion disk with a strong advective
component, namely the advective accretion disk, is hotter than the Keplerian disk
which explains successfully several systems e.g. Sgr $A^*$, GRS~1515+105 which
are expected to be consisting of hot disks. Therefore, such an accreting system may not
be of pure Keplerian type. At least close to the compact object it is an advective accretion disk.
In a sub-Keplerian disk, at far away off the black hole, the speed of the infalling matter is 
close to zero when it is practically out of the black hole's influence. However, temperature of the
system is finite resulting in finite sound speed at that radius. On the other hand, matter speed
at the black hole horizon reaches the speed of light ($c$), while, by causality condition, sound 
speed can be at the most $c/\sqrt{3}$. Therefore, there must be a location where matter speed
crosses local sound speed making the flow transonic. However, the second condition namely 
the inner boundary condition not necessarily be satisfied in the flow around a neutron star. Therefore
an accretion flow around a neutron star may or may not be transonic.

In the present article, I mainly concentrate on the sub-Keplerian accretion disk
with possible multitransonic flow.
If the flow is transonic, then the disk dynamics and corresponding outflow 
are influenced by the sonic locations/points \cite{c89,c96r,c99,m03,mg03}. Depending upon the flow energy (entropy) 
and angular momentum, a disk may exhibit one to four sonic points, through which all
however, matter may not pass. Sonic points in accretion disks play role
similar to the critical points in simple harmonic
oscillators, particularly to control the dynamics of the system. 
Indeed an accretion disk can be looked upon as a damped harmonic oscillator \cite{c90}. 

I organize the paper as follows. In the next section, I discuss accretion flow,
first the spherical one and then the disk flow, and formation of possible 
sonic points therein in the absence of any energy dissipation. In \S3, I analyze
how the rotation of the compact object affects the sonic points. Subsequently,
for completeness, I present the generalized set of accretion disk equations including 
possible dissipation effects in \S4, without going into their solutions.
Finally, I summarize in \S5.

\section{Accretion flow and formation of sonic points}
\label{accson}

First I discuss simple spherical accretion namely Bondi flow. Subsequently, I include
angular momentum into the equations set describing disk accretion. I show that while
the Bondi flow around a nonrotating compact star exhibits single sonic point the
disk flow may exhibit multiple sonic points.

As gravity is expected to be very strong close to the compact object, in principle I
should describe the flow system by the set of general relativistic equations. For a 
nonrotating compact object equations should be written in the Schwarzschild geometry
and for a rotating compact object they should be in the Kerr geometry. However,
that might hide transparency of the description,
as the full general relativistic set of equations is so cumbersome that
it is difficult to relate the terms with the physics they carry. Indeed under
some occasions a full general relativistic description is not required. Therefore,
I describe the system by pseudo-Newtonian approach. Here one uses Newtonian
set of equations only along with a modified gravitational force/potential such
that it mimics the general relativistic features approximately to describe
the system.

\subsection{Bondi flow}
\label{bondi}

\begin{figure}
\centering
\includegraphics[height=8cm]{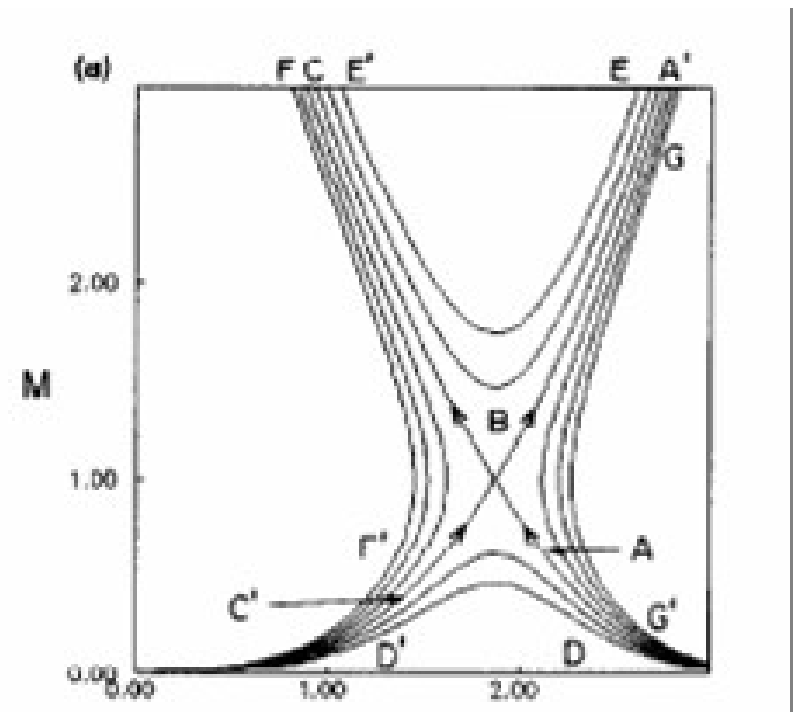}
\includegraphics[height=8cm]{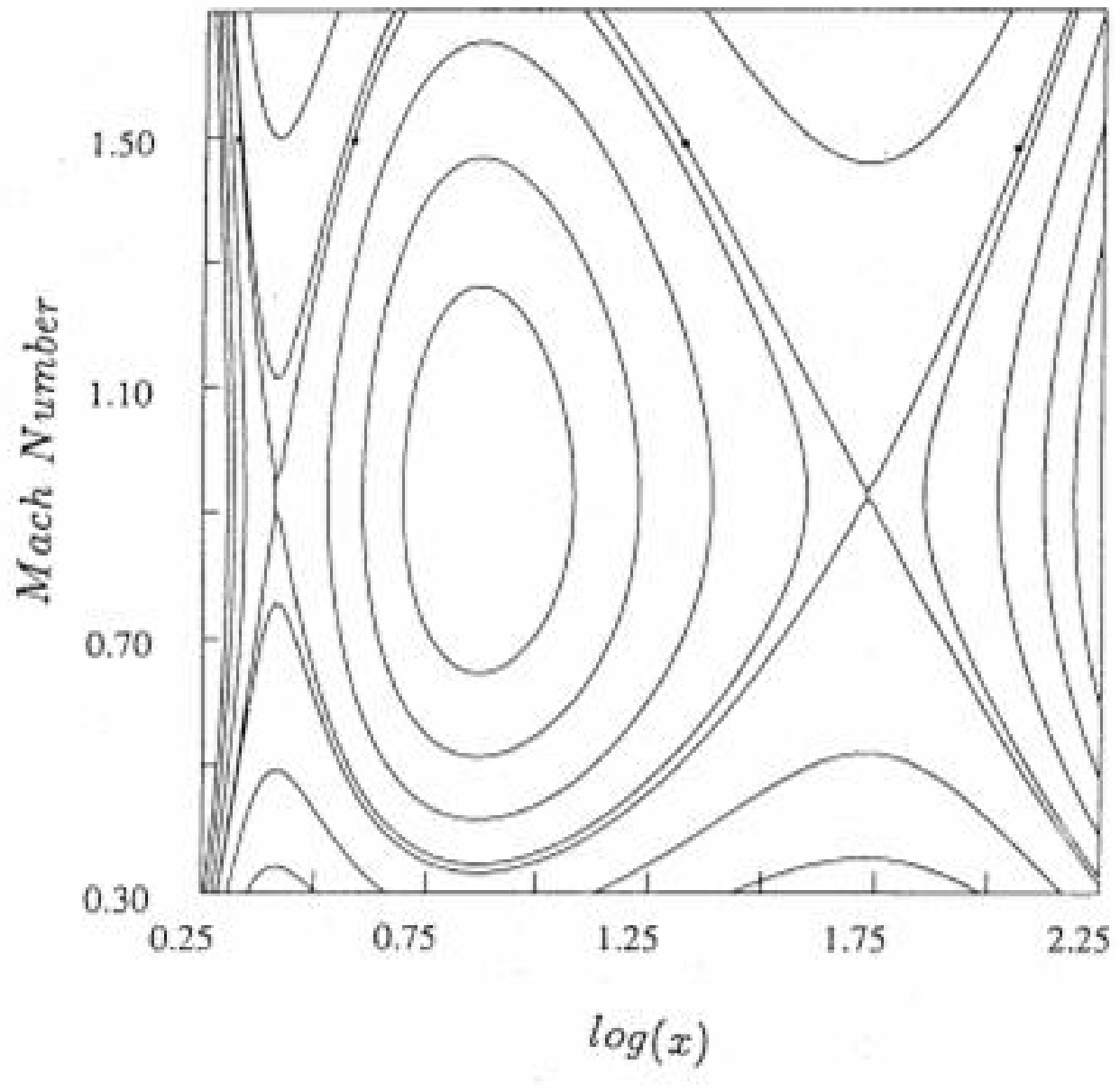}
%
%
\caption{Upper panel: Bondi flow: 
Mach number as a function of logarithmic radial coordinate for various energy of the flow.
The point `B' is sonic point and the solution marked by the arrow `A' indicates accretion
and `C$^\prime$' indicates wind. Lower panel: Disk flow: Same as Bondi flow; \cite{c90}.}
\label{figbond}       
\end{figure}

This happens for an isolated star when matter falls onto it from all directions,
resulting in a spherical accretion. Therefore, to describe the system, we consider
the spherical polar coordinate system where only nonzero component of velocity is $v_r=v$.
Hence the equations describing the steady-state flow with negligible
viscosity are given as:
\begin{eqnarray}
v\frac{dv}{dx}+\frac{1}{\rho}\frac{dP}{dx}+F_g=0,
\label{bonrmom}
\end{eqnarray}
\begin{eqnarray}
\frac{1}{x^2}\frac{d}{dx}\left(\rho v x^2\right)=0,{\rm\,\,\,hence\,\,\,}\rho v x^2={\rm constant}=\dot{M}_{ac},
\label{bonec}
\end{eqnarray}
where all the variables are expressed in dimensionless units.
$v$ is velocity in unit of light speed, $x=r/r_g$ where $r$ is radial coordinate and 
$r_g=GM_s/c^2$, $M_s,c,G$ are 
respectively mass of the central star, speed of light, Newton's gravitation constant, 
$P,\rho$ are respectively corresponding dimensionless pressure, density of the flow, 
$\dot{M}_{ac}$ is accretion rate and $F_g$ is gravitational force given by
\begin{eqnarray}
F_g&=&\frac{1}{x^2}:\,\,{\rm Newtonian},\\
&=&\frac{1}{(x-2)^2}:\,\,{\rm pseudo-Newtonian,\,\,Schwarzschild\,\,geometry}\,\cite{pw},\\
&=&\frac{(x^2-2j\sqrt{x}+j^2)^2}{x^3(\sqrt{x}(x-2)+j)^2}:\,\,{\rm pseudo-Newtonian,\,\,Kerr\,\,
geometry}\,\cite{m02},
\label{pseudopot}
\end{eqnarray}
where $j$ is specific angular momentum of the compact object.  
Integrating equation (\ref{bonrmom}) for an adiabatic flow to a nonrotating compact object I obtain
\begin{eqnarray}
E=\frac{1}{2}v^2+na^2-\frac{1}{x-2},
\label{bonber}
\end{eqnarray}
where polytropic index $n=1/(\gamma-1)$ and $\gamma$ is defined as $\gamma=a^2\rho/P$ with
$P=K\rho^\gamma$, $K$ is gas constant and $a$ is sound speed. If the Mach number of the flow is
defined as $M=v/a$, then at a constant energy $E$, it is clearly understood
from (\ref{bonber}) that $M-x$ trajectory is always hyperbolic-type \cite{bondi} with a sonic point
at $M=1$ shown in Fig. \ref{figbond} \cite{c90}. 
The solution marked `A' by an arrow indicates
accretion and marked `C$^\prime$' as wind. None of the other solutions are physical to describe
accretion and wind. However, for the flow around a rotating compact object, there might
be more than one sonic point but I will not discuss them here, this can be checked with a pseudo-Newtonian 
potential proposed for the Kerr geometry \cite{m02}.

\subsection{Disk accretion flow}

Now I consider a binary system when the compact object is closely associated with its
binary companion star and pulling matter off it. The gravitational force acts on the donor star
radially. On the other hand, the star is rotating and hence the matter detached off it
due to gravitational pull has angular momentum. As a result the matter infalls towards
the compact object in a spiral path forming a disk around it called the
accretion disk. Therefore, the radial momentum balance equation describing disk dynamics in 
steady-state in the absence of any energy dissipation is \cite{m03}
\begin{eqnarray}
v\frac{dv}{dx}+\frac{1}{\rho}\frac{dP}{dx}-\frac{\lambda^2}{x^3}+F_g=0,
\label{diskrmom}
\end{eqnarray}
where $\lambda$ is specific angular momentum of the infall which is a conserved
quantity in a nondissipative flow. The corresponding vertically averaged mass conservation reads as
\begin{eqnarray}
\frac{1}{x}\frac{d}{dx}\left(x\Sigma v\right)=0,\,\,{\rm hence}\,\,-4\pi x\Sigma v={\rm constant}=\dot{M}_{ac},
\label{diskec}
\end{eqnarray}
where $\Sigma\sim h(x)\rho$ is the (vertically integrated) column density \cite{matsum} with $\rho$ is 
density at the equatorial plane, $h(x)=a\sqrt{x/F_g}$ is the
half-thickness of the disk computed from the assumption of vertical equilibrium.
Now integrating (\ref{diskrmom}) for an adiabatic flow towards a nonrotating compact object 
I obtain specific energy of the flow
\begin{eqnarray}
E=\frac{1}{2}v^2+na^2+\frac{\lambda^2}{2x^2}-\frac{1}{x-2},
\label{diskber}
\end{eqnarray}
which is a conserved quantity for the nondissipative system. It is seen from (\ref{diskber}) that
at far away from the black hole and close to its event horizon located at $x=2$, last term (gravitational potential energy)
on the right hand side dominates over the last but one term (centrifugal energy) resulting
in a hyperbolic-type $M-x$ trajectory as shown in Fig. \ref{figbond} \cite{c96r,c90}. 
On the other hand, at an 
intermediate location, with an appropriate value of $\lambda$, centrifugal energy 
may dominate over gravitational potential energy resulting in an ellipse-type $M-x$ trajectory
as shown in Fig. \ref{figbond}.

\subsection{Formation of sonic points in disk accretion}

Combining (\ref{diskrmom}) and (\ref{diskec}) I obtain 
\begin{eqnarray}
\frac{dv}{dx}=\frac{\frac{\lambda^2}{x^3}-F_g+\frac{a^2}{\gamma+1}\left(\frac{3}{x}
-\frac{1}{F_g}\frac{dF_g}{dx}\right)}{v-\frac{2a^2}{(\gamma+1)v}}=\frac{N}{D}.
\label{dvdx}
\end{eqnarray}
At the sonic point $x=x_c$, $D=0$ and thus to have continuous $dv/dx$ at the same
location $N$ must vanish. Hence, applying $N=D=0$ at $x=x_c$ to (\ref{dvdx}), I obtain
velocity, sound speed, specific energy at $x_c$
\begin{eqnarray}
\nonumber
v_c&=&a_c\sqrt{\frac{2}{\gamma+1}},\,\,{\rm hence\,\,Mach\,\,number}\,\,M_c=\sqrt{\frac{2}{\gamma+1}},\\
\nonumber
a_c&=&\sqrt{(\gamma+1)\left(\frac{\lambda^2}{x_c^2}-F_{gc}\right)\left[\frac{1}{F_{gc}}
\left(\frac{dF_g}{dx}\right)_c-\frac{3}{x_c}\right]^{-1}},\\
E_c&=&\frac{2\gamma}{\gamma-1}\left[\frac{\frac{\lambda^2}{x_c^3}-F_{gc}}{\frac{1}{F_{gc}}
\left(\frac{dF_g}{dx}\right)_c-\frac{3}{x_c}}\right].
\label{sonqu}
\end{eqnarray}
I also compute a quantity that carries information of entropy 
\begin{equation}
\dot{\cal M}=(\gamma K)^n \dot{M}_{ac}
\label{sonen}
\end{equation}
which is a useful quantity to study the sonic point behavior. Now to obtain solutions for the accretion disk
one has to integrate (\ref{dvdx}) with a proper boundary condition. As our interest is in
transonic solutions, the matter necessarily passes through the sonic location and this 
I consider as our staring point of the integration. As the system under consideration is in the
steady-state, integrating from the sonic point to outwards (inwards) and vice versa are equivalent. However, $dv/dx$
given in (\ref{dvdx}) is of $0/0$ form. Therefore, applying l'Hospital's rule, I obtain
$dv/dx$ at $x=x_c$
\begin{equation}
\left(\frac{dv}{dx}\right)_c=-\frac{{\cal B}+\sqrt{{\cal B}^2-4{\cal A}{\cal C}}}{2{\cal A}},
\label{dvdxc}
\end{equation}
where,
\begin{eqnarray}
\nonumber
{\cal A}&=&1+\frac{2a_{c}^2}{(\gamma+1)v_c^2}+\frac{4a_{c}^2(\gamma-1)}{v_c^2(\gamma+1)^2},\\
\nonumber
{\cal B}&=&\frac{4a_{c}^2(\gamma-1)}{v_c(\gamma+1)^2}\left[\frac{3}{x_c}
-\frac{1}{F_{gc}}\left(\frac{dF_g}{dx}\right)_c\right],\\
\nonumber
{\cal C}&=&\frac{a_{c}^2}{\gamma+1}\left[\left(\frac{1}{F_{gc}}\left(\frac{dF_g}{dx}\right)_c\right)^2
-\frac{1}{F_{gc}}\left(\frac{d^2F_{g}}{dx^2}\right)_c-\frac{3}{x_c^2}\right]\\
&-&\frac{3\lambda^2}{x_c^4}-\left(\frac{dF}{dx}\right)_c
-\frac{a_{c}^2(\gamma-1)}{(\gamma+1)^2}\left[\frac{3}{x_c}-\frac{1}{F_{gc}}\left(\frac{dF_g}{dx}\right)_c
\right]^2.
\label{cof}
\end{eqnarray}
Now the value of discriminant ${\cal D}={\cal B}^2-4{\cal A}{\cal C}$ determines the type of
sonic point. Clearly ${\cal D}$ depends on the sonic location which can be computed
for a given $E_c$ from (\ref{sonqu}). In fact, the expression for $E_c$ can be rewritten as a fourth order 
polynomial of $x_c$. Hence, $E_c$ and $\lambda$ are the input parameters 
determining flow behavior for a given system.   

\begin{figure}
\centering
\includegraphics[height=3.5cm]{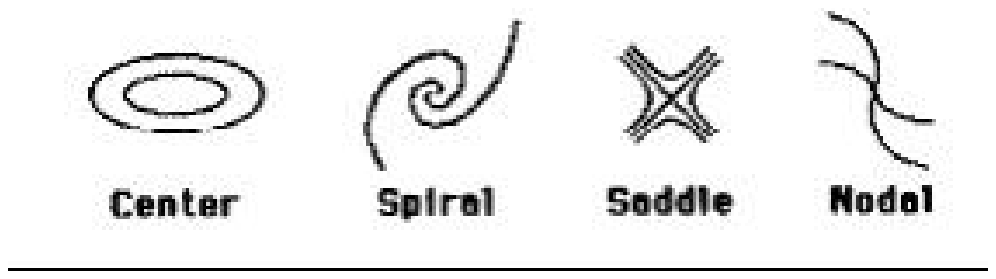}
\caption{Trajectories around various critical points where abscissa and ordinate are radial coordinate
and Mach number respectively; \cite{c90,st}.}
\label{figcrit}       
\end{figure}

There are four different types of sonic point shown in Fig. \ref{figcrit} \cite{st,c90}
classified according to the trajectory around it. 
(1) When ${\cal D}<0$ with ${\cal B}=0$, sonic point is center-type (elliptical trajectories in 
Fig. \ref{figbond}). (2) ${\cal D}<0$ with ${\cal B}\neq 0$ gives spiral-type sonic point.
For ${\cal D}\ge 0$ with (3) ${\cal AC}>0$ it is nodal-type, 
and (4) ${\cal AC}<0$, saddle-type (hyperbolic trajectories in Figs. \ref{figbond}). 
When ${\cal AC}=0$ for ${\cal D}\ge 0$, sonic points are called straight line.

\section{Sonic point analysis in disks around rotating compact objects}

\begin{figure}
\centering
\includegraphics[height=8cm]{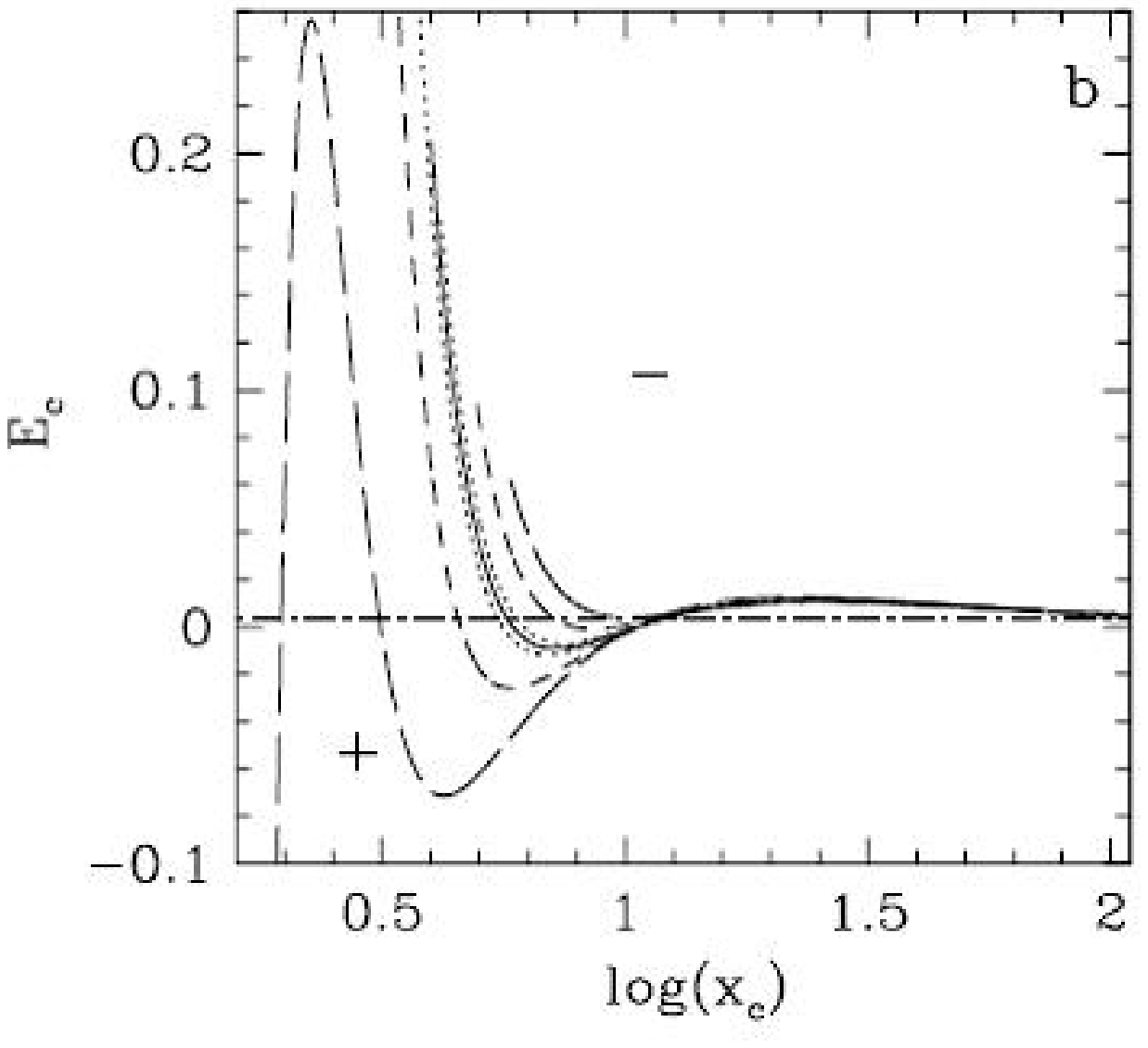}
\includegraphics[height=8cm]{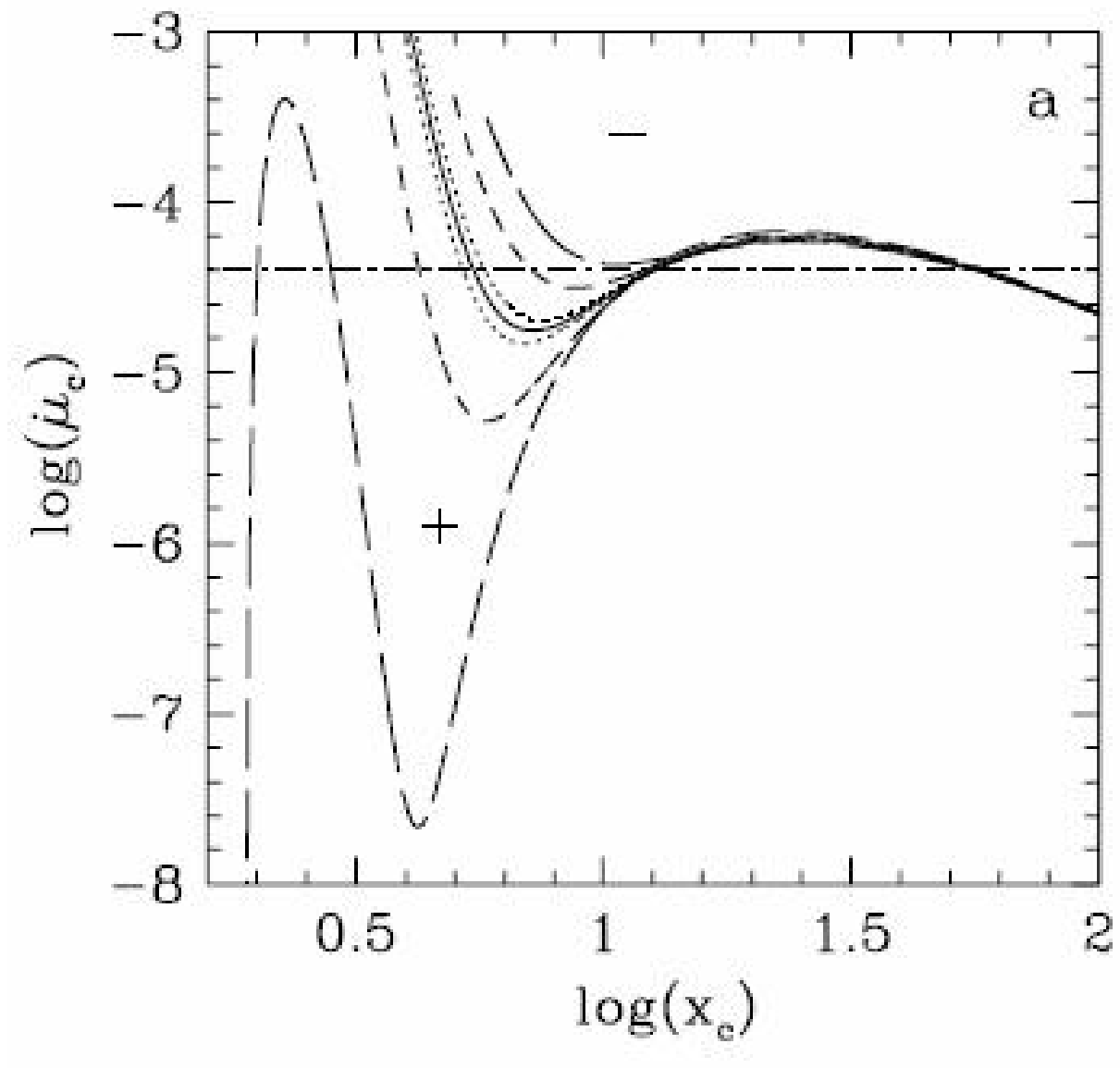}
\caption{Variation of energy (upper panel) and information of entropy (lower panel) as functions of logarithmic 
sonic radius when $j$ is a parameter. Central solid curve is for nonrotating compact object ($j=0$), 
while curves in regions of either side of it indicated by `$+$' and `$-$' are for prograde and and
retrograde orbits respectively. Different
curves from $j=0$ line to downwards are for $j=0.1,0.5,0.998$ and to upwards for $j=-0.1,-0.5,-0.998$.
The horizontal line indicates the curve of constant energy of $0.0065$ (upper panel) and of constant entropy 
of $5\times 10^{-5}$ (lower panel). Other parameters are $\lambda=3.3$, $\gamma=4/3$; \cite{m03}.}
\label{figson}       
\end{figure}

The energy and information of entropy at the sonic point $E_c$ and $\dot{\cal M}_c$ as functions of sonic location 
$x_c$ given by (\ref{sonqu}) and (\ref{sonen}) respectively
describe the loci of sonic point energy and entropy of the flow. Figure \ref{figson} \cite{m03}
shows their behavior for various choice of specific angular momentum of the compact object. 
As I consider a nondissipative disk, the energy and entropy remain conserved throughout.
The intersections of the curves 
by the constant energy and entropy line (which are the horizontal lines in the figure)
indicate the sonic points of the accretion disk for that particular energy and entropy and specific
angular momentum of the black hole (Kerr parameter). It is clearly seen that, 
at a particular energy and entropy, if the Kerr parameter
increases, the region where sonic points form
(as well as radii of marginally bound ($x_b$) and stable ($x_s$) orbits) shift(s) to a more
inner region of the disk and possibility to have all sonic points in the disk outside event horizon
increases. As an example, for $a=0.998$, the inner edge of the accretion disk
enlarges to such an extent that the fourth sonic point in the disk appears outside the event horizon.
On the other hand, for retrograde orbits (counter rotating cases), all the sonic points come close 
to each other tending to overlap for a particular value of the Kerr parameter (when $x_b$ and $x_s$ move to greater
radii). Therefore, as the value of the Kerr parameter decreases,
the possibility of forming individual sonic points decreases and thus shrinking the region
containing sonic points. This is very well-understood physically; as the
Kerr parameter decreases, total angular momentum of the system decreases resulting in
the disk tends to a {\it Bondi-like} flow, that has single saddle-type sonic point.

On the other hand, for a particular $a$, if $E_c$ and $\dot{\cal M}_{c}$ decrease, then
the possibility to have all the sonic points in the disk increases. This is understood from
Fig. \ref{figson} that as $E_c$ and $\dot{\cal M}_{c}$ decrease, the curve is more
likely to intersect the horizontal line. If $E_c=0.05$, then number of intersection
of all the curves, except the dashed one, by the $E_c=0.05$ line is one giving only
sonic point. The dashed curve with $a=0.998$ intersects twice. However, for 
$E_c=0.0065$ (as considered in Fig. \ref{figson}), the dashed curve intersects four times
exhibiting four possible sonic points in the flow and the other curves intersect thrice.
For a detailed description, see earlier works \cite{m03,mg03}.

Note that the sonic points occurring with a negative slope of the curve indicate the
locations of `saddle-type' sonic point and those with a positive slope indicate the `center-type' sonic point.
Thus the rotation of a black hole plays an important role in the formation of the sonic points
which are related to the structure of accretion disks and presumably formation of outflows and jets.

\section{Generalized set of equations describing a sub-Keplerian accretion disk}

So far, for simplicity, I have described disks without any dissipative energy. In principle
a disk must exhibit viscous dissipation making angular momentum varying with disk radius.
The hot disk flow with ion temperature $T\ge 10^9$K is also expected to generate significant 
nuclear energy mostly via proton-capture
reactions \cite{mc99,mc00}. On the other hand, significant energy is radiated out through
the inverse-Compton, bremsstrahlung and synchrotron effects, cooling the disk. In addition, 
significant energy is expected to be absorbed
through endothermic nuclear reactions, mostly via dissociation of elements \cite{mc99,mc00,mc01},
affecting disk dynamics significantly. Although the main purpose of the paper is to describe
the basic mechanism to form sonic points in accretion disks and for that the inviscid set of
equations without dissipation suffices, for completeness here I describe the generalized
set of equations including effects of energy dissipation. However, I will not go for the solutions 
of such equations set which is beyond the scope of present paper.

Therefore, in general one should incorporate two more equations namely angular momentum 
balance and energy equation apart from (\ref{diskrmom}) and (\ref{diskec}) as given below.
I express viscous dissipation $q^+$ in terms of shear stress
$W_{x\phi}$ as $q^+=W_{x\phi}^2/\eta$, where $\eta$ is the
coefficient of viscosity. Shakura \& Sunyaev \cite{ss73} parametrized shear stress in a Keplerian disk
by gas pressure with a constant $\alpha$, called Shakura-Sunyaev viscosity parameter, 
such that $W_{x\phi}=-\alpha P$. However, in an advective
disk, there must be a significant contribution due to ram pressure and thus shear stress may be
read as $W_{x\phi}=-\alpha(I_{n+1}P+I_n v^2\rho)h(x)$, where I consider vertically averaged disk
and $I_{n+1},I_n$ appear due to integration in the vertical direction \cite{matsum} when $P$ and $\rho$
are pressure and density respectively at the equatorial plane. However, if I use this expression of
shear stress in describing $q^+$, it loses information of actual shear, effect due to variation of
angular velocity in the disk. The proper way of writing shear stress should be 
$W_{x\phi}=\eta x \frac{d\Omega}{dx}$, where $\Omega$ is angular frequency of the flow.
If I substitute this into $q^+$, then final equation to be solved will contain a nonlinear derivative term
making it tedious to solve. Therefore, following Chakrabarti \cite{c96}, I express $q^+=-\alpha(I_{n+1}P+I_n v^2\rho)
h(x)x\frac{d\Omega}{dx}$, where one $W_{x\phi}$ in the square is expressed by pressure and other by 
actual shear.  Thus the angular momentum balance equation is
\begin{equation}
v\frac{d\lambda}{dx}=\frac{1}{\Sigma x}\frac{d}{dx}\left[x^2 q^+ \right].
\label{azmom}
\end{equation}
Then the energy equation may read as
   \begin{equation}
   \Sigma vT\frac{ds}{dx}=\frac{vh(x)}{\Gamma_3-1}\left(\frac{dP}{dx}-\Gamma_1\frac{P}{\rho}\frac{d\rho}{dx}\right)=Q^+-Q^-,
   \label{enteq}
   \end{equation}
  where $s$ is entropy density, $Q^+=q^+ +q_{\rm nuc}$ and $Q^-=q_{\rm ic}+q_{\rm br}+q_{\rm syn}$ when $q_{\rm nuc}$
is the nuclear energy released/absorbed in the disk and
$q_{\rm ic},q_{\rm br},q_{\rm syn}$ are respectively energy radiated out due to inverse-Compton, bremsstrahlung,
synchrotron effect. 
  Following Cox \& Giuli \cite{cg}, I define
  \begin{equation}
  \nonumber
   \Gamma_3=1+\frac{\Gamma_1-\beta}{4-3\beta},\\
   \nonumber
   \Gamma_1=\beta+\frac{(4-3\beta)^2(\gamma-1)}{\beta+12(\gamma-1)(1-\beta)},\\
   \beta=\frac{\rm gas\,\,pressure}{\rm total\,\,pressure}.
   \label{gambet}
   \end{equation}

To obtain generalized solutions of an accretion disk with a significant advective component, namely
a sub-Keplerian accretion disk, one has to solve equations (\ref{diskrmom}), (\ref{diskec}), 
(\ref{azmom}) and (\ref{enteq}) simultaneously along with the set of equations generating 
nuclear energy through the reactions taking place in a hot flow among all the isotopes
therein as given in earlier works in detail \cite{mc99,mc00,mc01}. 

\section{Summary}

I have described how the various types of sonic point form in an accretion disk. While sonic
points necessarily form in disks around a black hole, it need not be around a neutron star.
I have discussed that upto four sonic points may exist outside event horizon of a black hole
depending on the flow parameters and black hole's angular momentum. However, in a zero angular
momentum Bondi flow around a nonrotating compact object, there is only one sonic point.
When a disk has saddle and/or nodal -type sonic points, matter passes through them, while 
it does not for center and spiral -type sonic points. Therefore, the last two are not physical sonic points 
for accretion and wind because they may not be the part of solutions extending from infinity to
the black hole horizon. However, the corresponding branches
may be useful to explain certain scenarios in disks, as argued by some authors, e.g. formation
of shock helping to understand observed truncation of disks \cite{rao,rao2}, launching of outflows and jets \cite{c99}.
In these cases, as explained by previous authors \cite{c99,m03,mg03,c90,c96}, the infalling 
matter first becomes supersonic passing through the outer saddle-type sonic point, then the shock forms and
it becomes subsonic around a center-type sonic point, and finally passing through the inner saddle-type
sonic point it falls into the black hole. A detailed behavior of such solutions around a fast rotating
compact object should be investigated for understanding jet physics in future.


%
\index{paragraph}
%
%
%
%
%
%
%



%
%
%
%
%
%
%
\input{referenc}



\printindex
\end{document}

%% file: referenc.tex
%
%

%
%